\documentclass[
 reprint,
 superscriptaddress,   
 amsmath,amssymb,
 aps,
]{revtex4-2}

\usepackage{graphicx}
\usepackage{dcolumn}
\usepackage{bm}

\begin{document}

\title{Photon Momentum Enabled Symmetry Breaking and Nonlinear Photocurrents in the Centrosymmetric Dirac Semimetal PdTe}

\author{Sambhu G Nath}
\affiliation{Department of Physical Sciences, Indian Institute of Science Education and Research Kolkata, Mohanpur 741246, West Bengal, India}

\author{Subhadip Manna}
\affiliation{Department of Physical Sciences, Indian Institute of Science Education and Research Kolkata, Mohanpur 741246, West Bengal, India}

\author{R K Gopal}
\affiliation{Department of Physics and Material Science and Engineering, Jaypee Institute of Information Technology Noida, India}

\author{Chiranjib Mitra}
\email{Corresponding author: chiranjib@iiserkol.ac.in}
\affiliation{Department of Physical Sciences, Indian Institute of Science Education and Research Kolkata, Mohanpur 741246, West Bengal, India}

\date{\today}

\begin{abstract}

In centrosymmetric Dirac semimetals, second order nonlinear photocurrents are forbidden by the coexistence of time-reversal and inversion symmetries. Here, we demonstrate that finite photon momentum transfer acts as a dynamic symmetry breaking mechanism in PdTe, enabling nonlinear optical responses that are nominally forbidden in the centrosymmetric bulk. Through polarization sensitive measurements, we resolve distinct contributions from the circular photogalvanic effect (CPGE), geometric shift currents, and photon drag mediated processes. We show that the helicity dependent current vanishes at normal incidence and reverses sign with the angle of incidence, reflecting the coupling between photons and spin polarized surface states. Crucially, thickness dependent analysis reveals that the helicity dependent photocurrent component C scales with film thickness, establishing a robust bulk contribution enabled by momentum transfer. This confirms that incident photons provide the directional axis required to probe interband quantum geometry, rather than the response originating solely from surface states or strain. Our results demonstrate that optical excitation can dynamically reduce the effective symmetry of the system, enabling access to quantum geometric tensors and establishing PdTe as a promising platform for exploring nonequilibrium dynamics governed by photon momentum in high-symmetry topological materials.

\end{abstract}

\maketitle

\section{\label{sec:level1}Introduction}

Topological materials have attracted broad interest due to their unconventional electronic band structures, which give rise to symmetry protected surface states and robust transport properties arising from strong spin-orbit coupling and linear energy dispersion~\cite{hasan2010colloquium,qi2011topological}. Dirac semimetals (DSMs) form a key class within this family, hosting symmetry protected Dirac nodes that give rise to ultrahigh mobility, giant magnetoresistance, and other topological transport phenomena~\cite{liu2014stable,neupane2014observation,he2014quantum,liang2015ultrahigh}. In ideal DSMs, the coexistence of time-reversal and inversion symmetries enforces the degeneracy of Weyl nodes with opposite chirality at identical momenta in the Brillouin zone, resulting in the cancelation of net Berry curvature and suppression of spin-polarized transport~\cite{armitage2018weyl,young2012dirac}. Consequently, direct spin generation, electrical manipulation of spin polarization, and helicity dependent optical responses are prohibited by symmetry in pristine Dirac semimetals.

Nonlinear optical probes offer a powerful route to uncover symmetry breaking and Berry curvature effects that remain hidden in linear transport. In particular, the bulk photovoltaic effect (BPVE), an intrinsic second-order nonlinear response that converts optical excitation into a dc electrical current without relying on built-in electric fields or conventional p-n junctions is a key tool for probing such phenomena \cite{morimoto2016topological,ahn2020low}. However, in perfectly inversion symmetric Dirac semimetals (DSMs), the BPVE is strictly forbidden, because the second-order conductivity tensor $\sigma_{ijk}$ must vanish in any centrosymmetric crystal, thereby eliminating all second-order photocurrent responses~\cite{sodemann2015quantum}. To overcome this constraint, inversion symmetry can be relaxed through strain engineering, while magnetic doping breaks time-reversal symmetry and by circularly polarized excitation selectively addresses chiral electronic states, together enabling asymmetric optical transitions and finite nonlinear photocurrents~\cite{ahn2020low,liang2022strain,chen2024defect}. Recent experiments on strain free Cd$_3$As$_2$ nanostructures have revealed helicity dependent photocurrents even in nominally centrosymmetric systems, indicating that optical fields can dynamically break symmetry and activate spin-polarized photocurrent generation \cite{wang2023spatially}. Similar mechanisms have been proposed in two-dimensional Dirac systems such as graphene, where substrate induced inversion asymmetry or interface fields enable circular and linear photogalvanic effects (LPGE)~\cite{glazov2014high,ganichev2003spin}. These observations collectively indicate that DSMs can host substantial nonlinear bulk photocurrents once even subtle deviations from perfect inversion symmetry, static or optically induced, are present.

Among the BPVE mechanisms, the shift and injection currents are well known second-order response that requires broken inversion symmetry. Microscopically, the shift current originates from real space displacements of Wannier centers associated with the initial and final electronic states during an optical transition. In contrast, the injection current arises from the asymmetric population of carriers in momentum space, leading to a net velocity imbalance immediately after photoexcitation \cite{tan2016shift,dai2023recent}. Observation of these photocurrents, however, necessitates a reduction of crystal symmetry. In particular, circular shift (CS) and linear injection (LI) currents change sign under both time-reversal ($\mathcal{T}$) and inversion ($\mathcal{P}$) operations, and are therefore generally expected to appear only in materials that break parity, such as magnetic or antiferromagnetic systems \cite{wang2020electrically}. However, recent theoretical advances have clarified how momentum transfer from incident photons, the photon drag effect, can restore nonlinear photocurrents in systems where they are usually forbidden. Xiong \textit{et al}. showed that polariton driven momentum transfer can generate quantum geometric photocurrents in high symmetry materials where conventional nonlinearities vanish \cite{xiong2022polariton}. Shi \textit{et al}. further demonstrated that finite shift currents can appear in centrosymmetric crystals when nonvertical, momentum transferring transitions, enabled by photon or polariton drag, activate an intrinsic shift current dipole set by the interband quantum geometry \cite{shi2021geometric}. More recently, Xie \textit{et al}. showed that photon drag induced bulk photovoltaic responses can be written directly in terms of the quantum geometric tensors, revealing that these momentum assisted photocurrents originate from quantities such as the quantum metric and Berry connection that govern nonvertical transitions even in centrosymmetric systems \cite{xie2025photon}.

Here we investigate polarization dependent photocurrents in the centrosymmetric Dirac semimetal PdTe. Using circularly and linearly polarized excitation, we resolve distinct contributions from the circular photogalvanic effect (CPGE), geometric shift current, surface photogalvanic effect (SPGE), and other photon drag enabled photocurrents. The emergence of helicity and polarization sensitive signals in PdTe reveals unconventional symmetry breaking and nonlinear optical pathways in a system where such effects are nominally forbidden. As discussed below, the observation of a strong polarization dependent photocurrent in PdTe reflects a fundamental momentum transfer process, whereby the incident photons supply a directional axis that effectively lowers the symmetry of the system. In this picture, the polarization state of light controls how this injected momentum is converted into charge flow, thereby setting the photocurrent direction. Our findings reveal that photon drag provides a general route to activating directional dc optical nonlinearities in centrosymmetric and polycrystalline Dirac semimetals, even in the absence of intrinsic inversion symmetry breaking. Our results establish PdTe as a promising platform for exploring nonequilibrium spin and charge dynamics in Dirac materials, and demonstrate how nonlinear phototransport measurements can reveal otherwise hidden topological characteristics and symmetry induced effects in centrosymmetric crystals.

\section{\label{sec:level1}EXPERIMENTAL METHODS}
Thin films of the Dirac semimetal PdTe with varying nominal thicknesses were deposited on single-crystal Al$_2$O$_3$ substrates using pulsed laser deposition (PLD). The PLD technique, owing to its highly nonequilibrium growth conditions, enables reliable transfer of target stoichiometry to the deposited films. High purity Pd and Te precursor materials were used to prepare the stoichiometric PdTe target. A KrF excimer laser operating at a wavelength of 248~nm was employed with a repetition rate of 4~Hz. Prior to deposition, the growth chamber was evacuated to a base pressure of $6 \times 10^{-6}$~mbar, following which an argon ambient was introduced and maintained at a partial pressure of $5 \times 10^{-1}$~mbar during deposition. The substrates were heated to $260^\circ$C, and the laser fluence was set to approximately 1.2~J/cm$^2$ to ensure stable ablation and uniform film growth. After deposition, the PdTe films were annealed \emph{in situ} at the growth temperature for 15~minutes under base pressure to improve crystallinity and reduce surface roughness.

\begin{figure}
\centering
\includegraphics[scale=0.28]{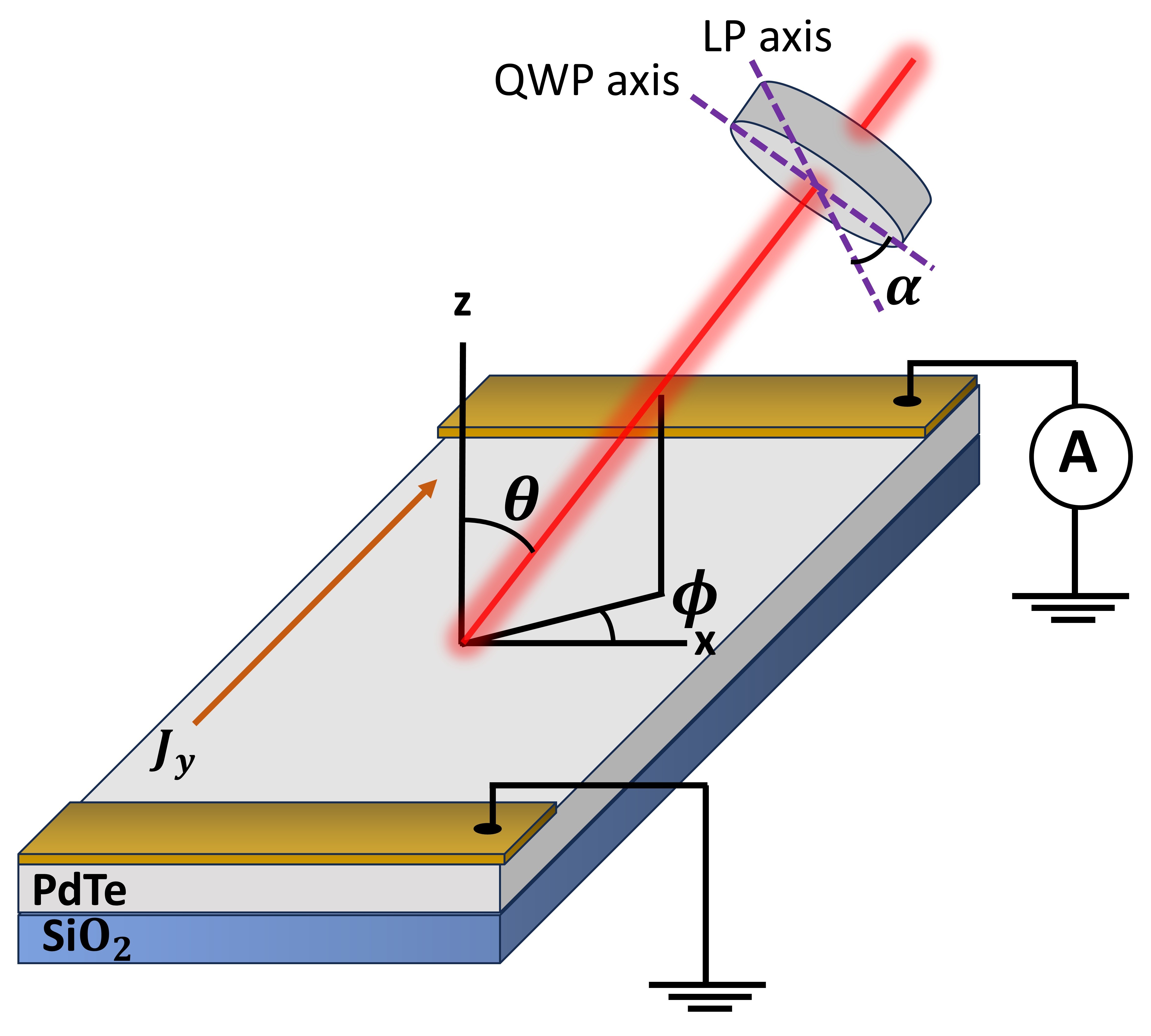}
\caption{\label{fig:11} Schematic illustration of the experimental geometry, showing the directions of the incident laser beam and the measured photocurrent with respect to the sample normal.}
\end{figure}

\section{\label{sec:level1}Results and Discussion}

\begin{figure*}
\centering
\includegraphics[scale=0.60]{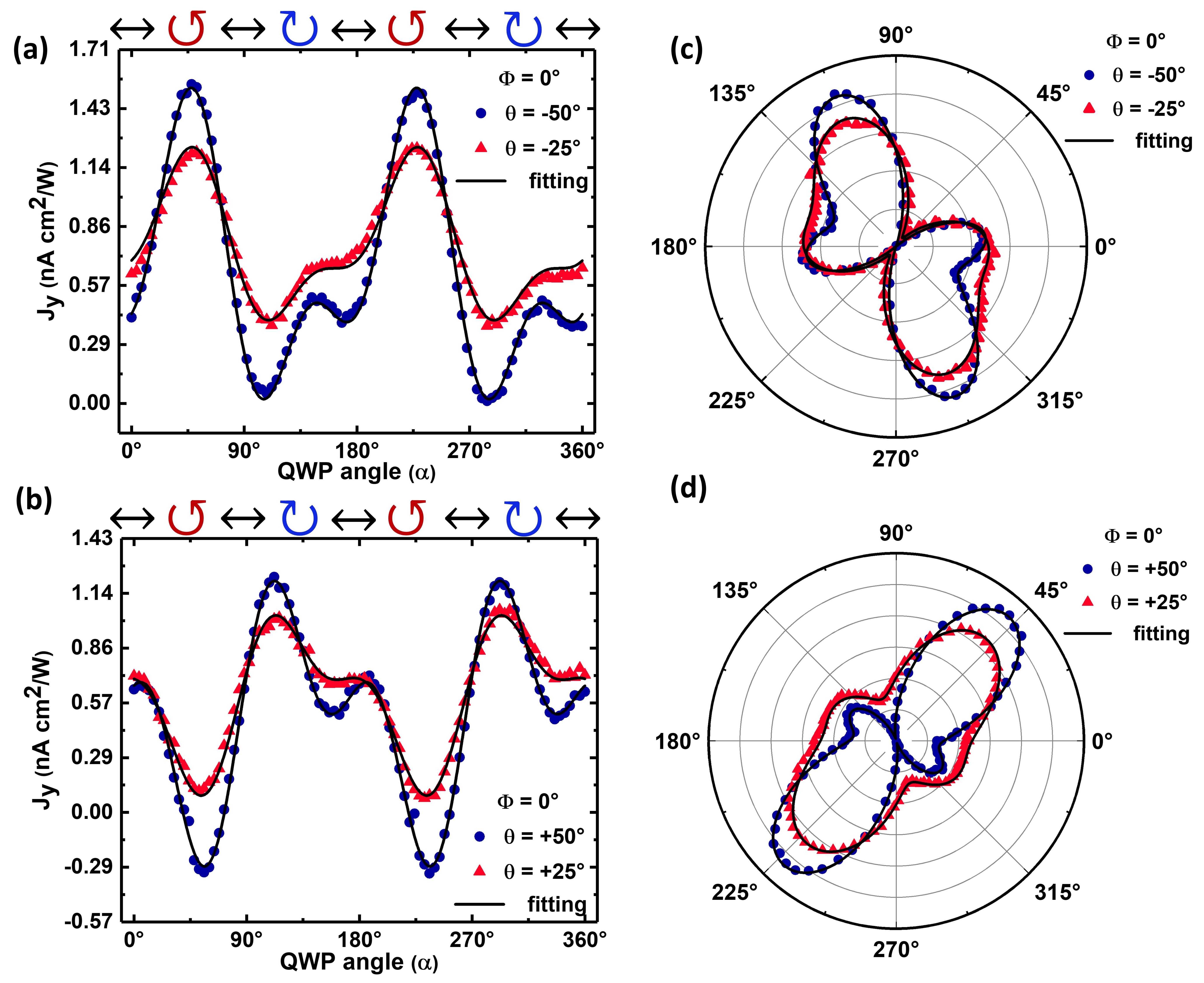}
\caption{\label{fig:1}(a,b) Photocurrent density $J_y$, normalized by the incident light intensity, as a function of the quarter-wave plate rotation angle $\alpha$ for selected negative (a) and positive (b) angles of incidence $\theta$ relative to the sample normal. The corresponding polarization states of the incident light for different values of $\alpha$ are indicated on the top axis. (c,d) Polar plots of the same data shown in panels (c) and (d), respectively. Solid lines represent fits to Eq.(\ref{qwp}), as discussed in the main text.}
\end{figure*}

\begin{figure*}
\centering
\includegraphics[scale=0.6]{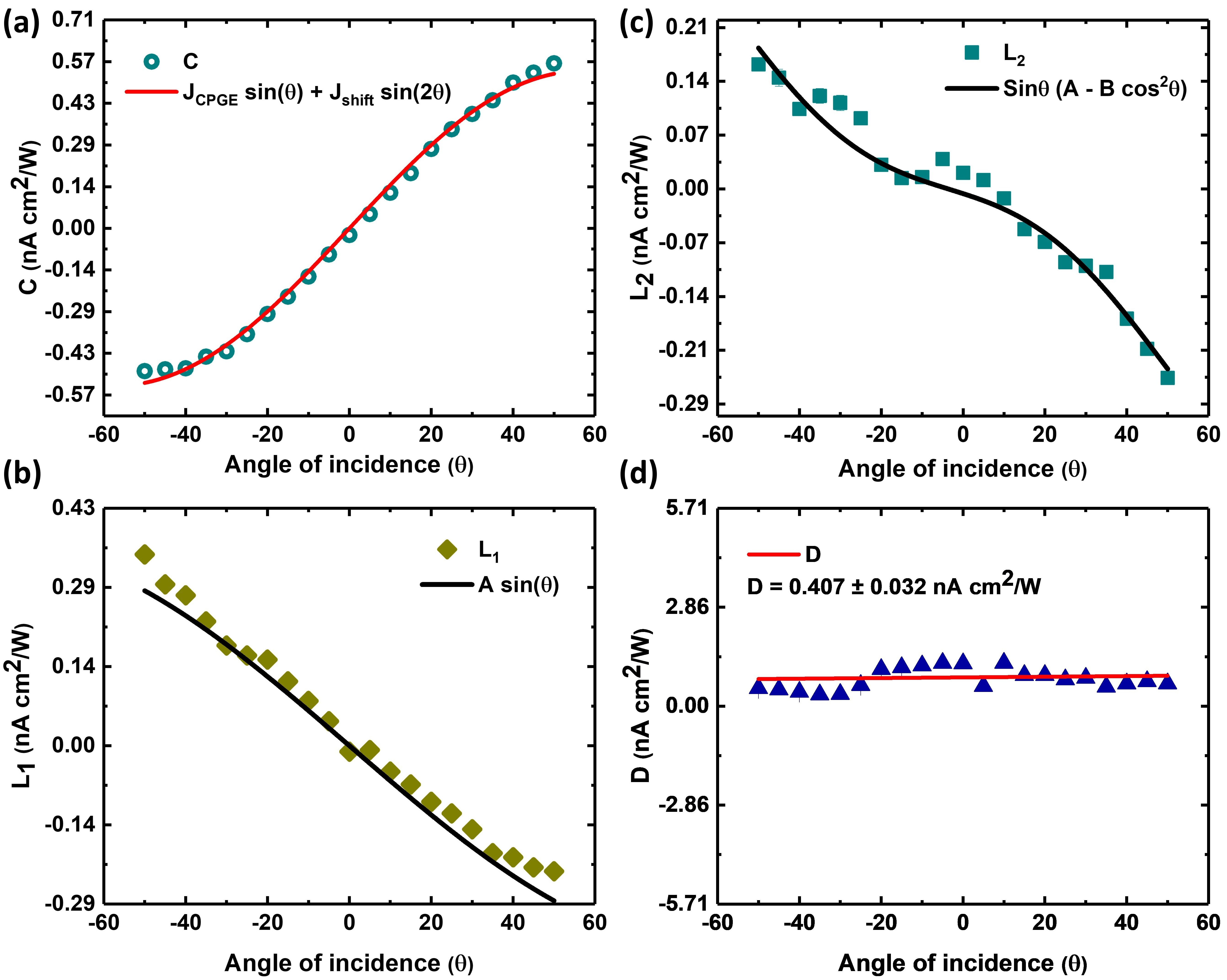}
\caption{\label{fig:3}Angle of incidence ($\theta$) dependence of (a) $C$, (b) $L_1$, (c) $L_2$, and (d) $D$. 
The solid curves denote fits to the corresponding functional forms indicated in each panel, with the extracted fitting parameters shown alongside.}
\end{figure*}

In our measurements, a 691~nm continuous wave laser beam was focused to a spot size of $\sim 100~\mu$m on unbiased PdTe devices, and the resulting transverse photocurrent $J_y$ was measured between the edges using a lock-in amplifier (SR830). The incident light intensity was modulated at a reference frequency of $227~\text{Hz}$ using a mechanical chopper for the lock-in amplifier measurements. The measured photocurrent generally contains two primary contributions, a photovoltaic (PV) component and a photothermoelectric (PT) component. To suppress the thermoelectric background arising from laser induced temperature gradients, the excitation beam was focused at the center of the device. At this position, the photocurrent remains finite but reverses its polarity when the laser spot is moved toward either contact, indicating the presence of a thermoelectric contribution. By positioning the beam at the device center, the sample is heated symmetrically, thereby minimizing lateral temperature gradients and reducing the PT background. The polarization dependent photocurrent was investigated by controlling the polarization state of the incident light with a quarter wave plate (QWP) rotated about the beam propagation axis. The QWP rotation angle $\alpha$ is defined to be zero when its fast axis lies in the $x$-$z$ plane, where the light is initially linearly $P$-polarized. As the QWP is rotated, the polarization of the incident light evolves continuously, cycling from linear $P$-polarization ($\alpha = 0^\circ$) to left circular polarization (LCP, $\alpha = 45^\circ$), back to linear $P$-polarization ($\alpha = 90^\circ$), then to right circular polarization (RCP, $\alpha = 135^\circ$), and finally returning to linear $P$-polarization ($\alpha = 180^\circ$). In this configuration, the circular polarization states repeat with a $180^\circ$ period, while the linear $P$-polarization repeats every $90^\circ$. By recording $j_y$ as a function of the QWP angle $\alpha$ for different incident angles $\theta$ of the excitation beam, the polarization resolved photocurrent contributions in PdTe were mapped systematically. The incident angle ($\theta$) of the excitation beam with respect to the sample normal was varied from $+50^{\circ}$ to $-50^{\circ}$ with a precision of $0.5^{\circ}$, thereby tuning the in-plane component of the incident optical wave vector. A schematic representation of the experimental setup employed for polarization dependent photocurrent measurements is shown in Fig.\ref{fig:1}(a).
The photocurrent traces obtained for several positive and negative incident angles $\theta$ (corresponding to illumination from opposite sides of the sample normal) are shown in the Cartesian representations in Fig.\ref{fig:1}(b,c). For all values of $\theta$, the current exhibits a periodic dependence on the polarization angle $\alpha$ with a characteristic $180^{\circ}$ periodicity. The polarization states associated with different values of $\alpha$ are indicated along the top axis of Fig.\ref{fig:1}(b,c). The modulation amplitude of $J_y$ is observed to increase with the magnitude of $\theta$, and the photocurrent profiles for $+\theta$ and $-\theta$ show clear asymmetries, indicating different polarization responses when the direction of incidence is reversed. The same datasets, when represented in polar coordinates in Fig.\ref{fig:1}(d,e) provide an alternative visualization. The resulting two-lobed asymmetric patterns highlight the $180^{\circ}$ periodicity through their distinct twofold symmetry. The variation between the responses for positive and negative incident angles manifests as a rotation or flip of the two-lobed asymmetric shaped contours.

When the excitation beam is incident obliquely in the $x$--$z$ plane, the transverse photocurrent $J_{y}$ exhibits a pronounced dependence on the polarization angle. The polarization angle ($\alpha$) dependence of the photocurrent at different angles of incidence is well described by the following expression (Eq.\ref{qwp}), from which the modulation amplitudes are extracted. 
\begin{equation}
J_y = C \sin(2\alpha) + L_1 \sin(4\alpha) + L_2 \cos(4\alpha) + D .
\label{qwp}
\end{equation}

The coefficient $C$ captures the helicity dependent component of the photocurrent, as rotation of the quarter wave plate continuously tunes the polarization between left  and right circular states, giving rise to the characteristic $\sin(2\alpha)$ dependence. The remaining coefficients in Eq.~(1) describe helicity independent responses. $L_1$ and $L_2$ correspond to components that vary with the linear polarization of the incident light, while $D$ accounts for a polarization insensitive background. The amplitudes of these four contributions extracted from the fits are plotted as functions of the incident angle $\theta$, as shown in in Fig(\ref{fig:3}). Their evolution with $\theta$ is analyzed to understand the underlying physical mechanisms and microscopic origins of each component.

\begin{figure*}
\centering
\includegraphics[width=\textwidth]{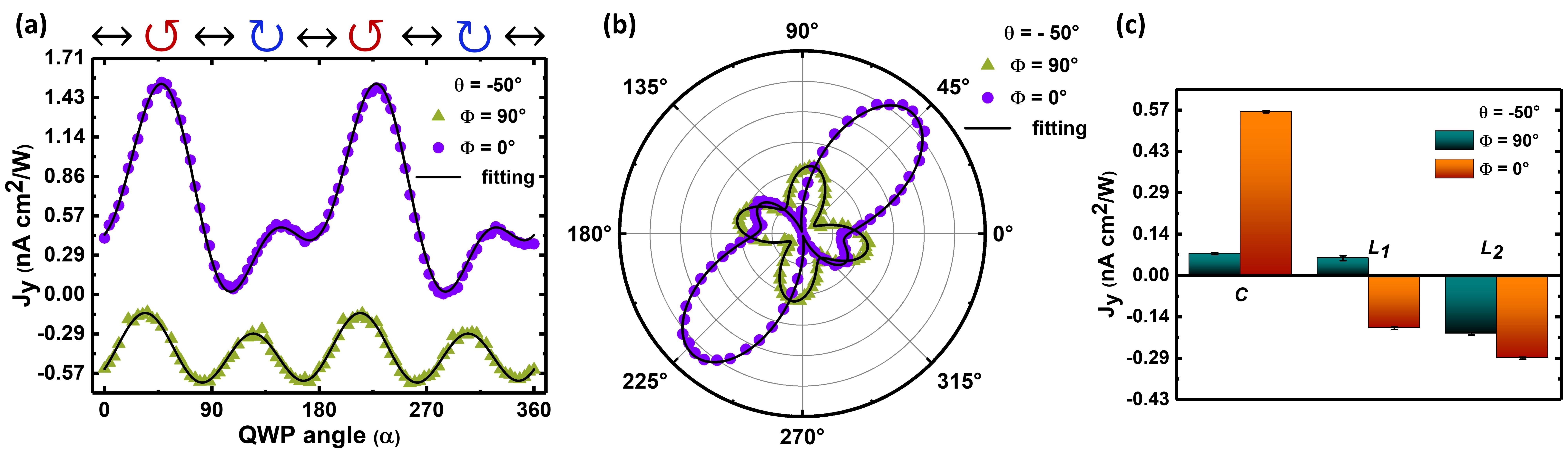}
\caption{\label{fig:2}(a) Polarization dependent modulation of the photocurrent measured at different values of $\phi$ for a fixed angle of incidence $\theta = -50^\circ$.
(b) Polar representation of the same data shown in panel (a). Solid lines represent fits to Eq.(\ref{qwp}), as discussed in the main text. (c) Comparison of the extracted photocurrent components $C$, $L_1$, and $L_2$ of $J_y$ for two different values of $\phi$.}
\end{figure*}

The helicity dependent photocurrent component $C$ shows a pronounced sensitivity to the incident angle $\theta$, as illustrated in Fig.\ref{fig:3}(a). Its magnitude is minimal at normal incidence, increases as magnitude of $\theta$ increases, and reverses sign when the light is incident from the opposite side of the surface normal. This behavior indicates that $C$ can have contribution primarily from spin-polarized surface states, such as the Fermi arc states expected in PdTe \cite{yang2023coexistence,chapai2023evidence}. Consistent with this interpretation, recent experiments on strain free Cd$_3$As$_2$ nanostructures have demonstrated that helicity dependent photocurrents can arise from surface state dominated transport \cite{wang2023spatially}. If the helicity driven photocurrent indeed originates from the surface states, its dependence on the plane of incidence should mirror the underlying spin texture \cite{mciver2012control}. By comparing the magnitude of $C$ for different planes of incidence (with the $x$-$z$ plane corresponding to $\phi = 0^\circ$), we can map how the surface state spin distribution projects onto the measured current direction. Since surface state contributions are tied to the spin texture, measuring the variation of $C$ as a function of $\phi$ provides a direct probe of how the spin distribution contributes to the total current, helping to distinguish these surface effects from bulk mediated photon drag processes. The effect can be understood in terms of the circular photogalvanic effect (CPGE) \cite{ganichev2003spin}, which originates from an asymmetric spin polarization created by selective optical excitation of spin–momentum locked surface states, where inversion symmetry is broken \cite{mciver2012control,pan2017helicity,roy2022photothermal}. In PdTe, the bulk remains centrosymmetric and does not contribute to CPGE, while the surface exhibits local inversion symmetry breaking \cite{yang2023coexistence}. As the incidence angle increases, the in-plane component of the photon angular momentum increases, enhancing the coupling to the surface in-plane spin angular momentum. Due to spin-momentum locking, this spin selective excitation produces spin-polarized carriers via spin imbalance, leading to a helicity dependent photocurrent that increases with $\theta$. When the light is incident from the opposite side, the orientation of the excited surface spins reverses, resulting in a corresponding sign change of $C$. For the $y$–$z$ plane of incidence ($\phi = 90^\circ$), we observe nearly an order of magnitude suppression of the helicity dependent component $C$ Fig.\ref{fig:2}(a-c). This behavior is consistent with the presence of spin-momentum locked surface states, as the helicity driven photocurrent is generated transverse to the light scattering plane. Theoretical predictions and experimental observations have confirmed CPGE signals originating from Fermi arc surface states in Weyl semimetals like RhSi \cite{chang2020unconventional,rees2021direct}. A analogous process can arise in Dirac semimetals via their topological surface states. Under equilibrium conditions, the paired Fermi arcs on the surfaces exhibit opposite spin polarizations, yielding no net charge current. In thick Dirac semimetal films, though, the top and bottom surfaces become effectively isolated, with photoabsorption largely confined to the illuminated side. Circularly polarized illumination thus preferentially populates carriers from a single Fermi arc branch, inducing a transient surface spin imbalance. Spin-momentum locking then transforms this imbalance into a unidirectional charge current, enabling a measurable CPGE \cite{wang2023spatially}. From the symmetry considerations of the material, the CPGE response is expected to follow a $\sin\theta$ dependence, arising from the projection of the photon angular momentum onto the surface where inversion symmetry is broken.

Another possible contribution to $C$ arises from the photon drag bulk photovoltaic effect, which can generate helicity dependent responses even in centrosymmetric materials through momentum transfer from the incident light. Xie \textit{et al.} have theoretically demonstrated that these momentum assisted excitations enable a shift current response even in nonmagnetic, centrosymmetric materials, where conventional shift currents are otherwise forbidden \cite{xie2025photon}. Importantly, the resulting photon drag induced shift current exhibits a characteristic polarization dependence proportional to $\sin(2\alpha)$, reflecting its helicity odd nature and the symmetry constraints imposed by the finite photon momentum. Moreover, they predict that this photon drag induced shift current is oriented transversely to the plane of incidence, providing a clear symmetry signature that can be distinguished by the dependence of $C$ on the plane of incidence. In addition, its magnitude increases with  $\theta$ because the in-plane component of the photon linear momentum, $q_{\parallel}=q\sin\theta$, becomes larger for more oblique illumination. Thus, the photon drag induced circular shift current can naturally account for the observed \(C\), as the transfer of photon momentum enables circularly polarized light to generate a directional charge response even in the absence of intrinsic inversion symmetry breaking. From symmetry considerations, the helicity dependent photon drag induced shift current is expected to exhibit a $\sin(2\theta)$ dependence, arising from the interplay between the in-plane photon momentum and the symmetry allowed nonlinear response tensor. This is fully consistent with our experimental observations, where $C$ exhibits both $\sin\theta$ and $\sin(2\theta)$ contributions, providing clear evidence for the coexistence of CPGE and the circular photon drag induced shift current, as shown in Fig.\ref{fig:3}(a).

\begin{figure*}
\centering
\includegraphics[width=\textwidth]{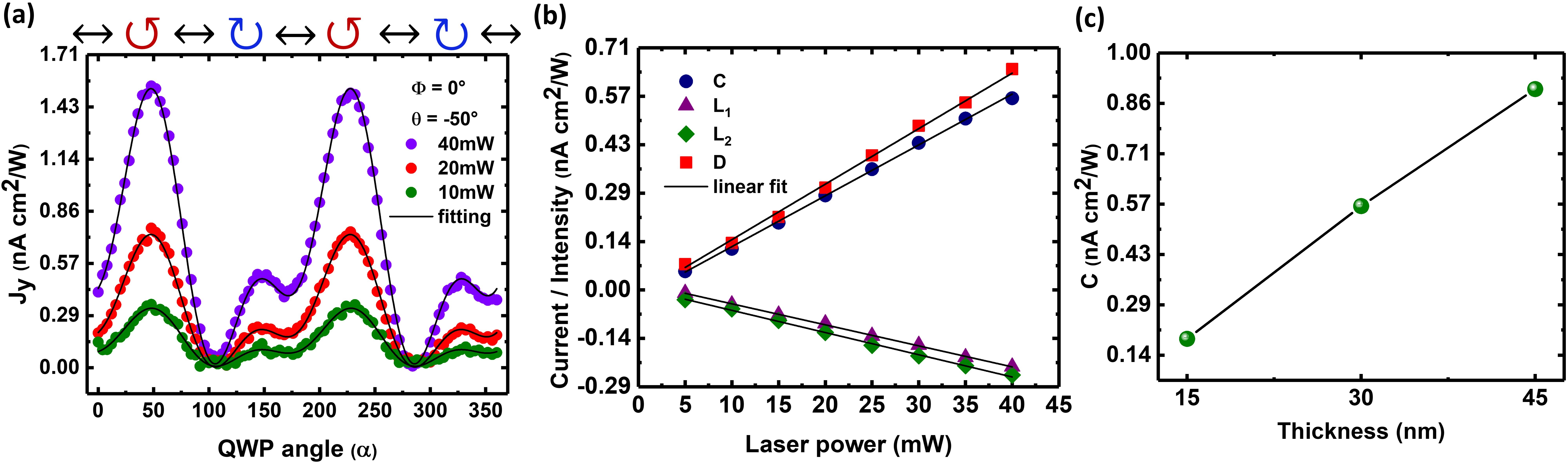}
\caption{\label{fig:4} (a) Photocurrent $J_y$, measured at different incident laser powers, plotted as a function of the quarter-wave plate (QWP) rotation angle $\alpha$. (b)Dependence of the photocurrent components $C$, $L_1$, $L_2$, and $D$ on the average incident laser power. All components exhibit linear behavior in the power range of 0-40~mW, as indicated by the solid-line fits. (c) Thickness dependence of the helicity dependent component of photocurrent (C).}
\end{figure*}

Helicity dependent photocurrent can also have a possible origin from  strain induced lifting of band degeneracy in otherwise centrosymmetric bulk states. A similar mechanism has been demonstrated in strained Cd$_3$As$_2$ thin films grown on GaAs(111)B substrates, where epitaxial strain produces spin splitting in the otherwise degenerate bulk bands, enabling a finite circular photogalvanic effect (CPGE) along with a possible origin from topological surface states \cite{liang2022strain}. In that system, the CPGE amplitude was found to decrease systematically with increasing film thickness, consistent with the gradual relaxation of strain away from the substrate interface. In contrast, our PdTe films exhibit the opposite trend. The magnitude of $C$ increases with thickness as shown in Fig.\ref{fig:4}(c). This behavior indicates that the helicity dependent photocurrent in PdTe is unlikely to originate from strain induced spin splitting of trivial bulk bands, and instead points toward surface state related or photon drag induced mechanisms that remain robust as the film becomes thicker. 

The linear polarization dependent components $L_1$ and $L_2$ exhibit a pronounced dependence on the angle of incidence $\theta$, as shown in Fig.\ref{fig:3}(b,c). Both components display odd symmetry with respect to $\theta$, their magnitudes are strongly suppressed at normal incidence ($\theta = 0$) and increase with increasing $\theta$. Furthermore, both $L_1$ and $L_2$ reverse sign upon changing the sign of $\theta$, taking opposite values for positive and negative incident angles. 

Xie \textit{et al.} have theoretically demonstrated that the linear polarization dependent photocurrent components $L_1$ and $L_2$ correspond to injection and Fermi surface contributions \cite{xie2025photon}. In our measurements, the observed behavior of $L_1$ further suggests an origin associated with the surface photogalvanic effect (SPGE)\cite{gurevich1993photomagnetism}. The SPGE arises from an anisotropic momentum distribution of photoexcited carriers in the near surface region, generated by interband optical transitions and enhanced by diffuse surface scattering \cite{mikheev2018interplay}. For light incident, the interband transition probability along the plane of incidence contains a term proportional to $(\mathbf{E}\cdot\mathbf{K})^2$, where $\mathbf{E}$ denotes the electric field component in the plane of incidence and $\mathbf{K}$ is the quasimomentum of the photogenerated electrons. Upon varying the polarization of the incident light, an additional electric field component perpendicular to the plane of incidence is introduced, giving rise to an extra contribution to the transition probability of the form $\sim 2\mathrm{Re}(E_x^{*}E_y^{*} K_x K_y)$. In the presence of this out-of-plane field component, theoretical analyses predict that the transverse SPGE current exhibits a $\sin\theta$ dependence on the incidence angle \cite{gurevich1993photomagnetism,mikheev2018interplay}. Consistent with this prediction, we observe that the experimentally extracted $L_1$ component follows a clear $\sin\theta$ variation with $\theta$, as shown in Fig.\ref{fig:3}(b).

The dependence of $L_2$ on incident angle suggests that it originates from photon drag related processes, specifically the linear injection current and the linear photon drag effect (LPDE) \cite{xie2025photon}. As discussed above, $L_2$ exhibits an odd dependence on the angle of incidence $\theta$, being strongly suppressed at normal incidence and increasing in magnitude with increasing $\theta$, while reversing sign for opposite values of $\theta$, as shown in Fig.\ref{fig:3}(c). Such behavior is characteristic of photon drag driven photocurrents, which rely on the transfer of the in-plane component of photon momentum to charge carriers. Theoretical analyses by Xie \textit{et al.} have shown that, in centrosymmetric systems, photon drag can enable linear injection currents by imparting a finite crystal momentum during optical excitation, thereby lifting the symmetry constraints that otherwise forbid such responses \cite{xie2025photon}. In this framework, the resulting current scales with the in-plane photon momentum and is therefore expected to vary as $\sin\theta(A-B\cos^2\theta)$, consistent with the experimentally observed angular dependence of $L_2$. Microscopically, both the linear injection and LPDE currents arise from an asymmetric carrier population and acceleration induced by momentum transfer from the electromagnetic field, rather than from inversion symmetry breaking in the bulk. The agreement between the observed $\theta$ dependence of $L_2$ and the theoretical predictions supports a photon drag mediated origin for this component.

\begin{figure*}
\centering
\includegraphics[scale=0.6]{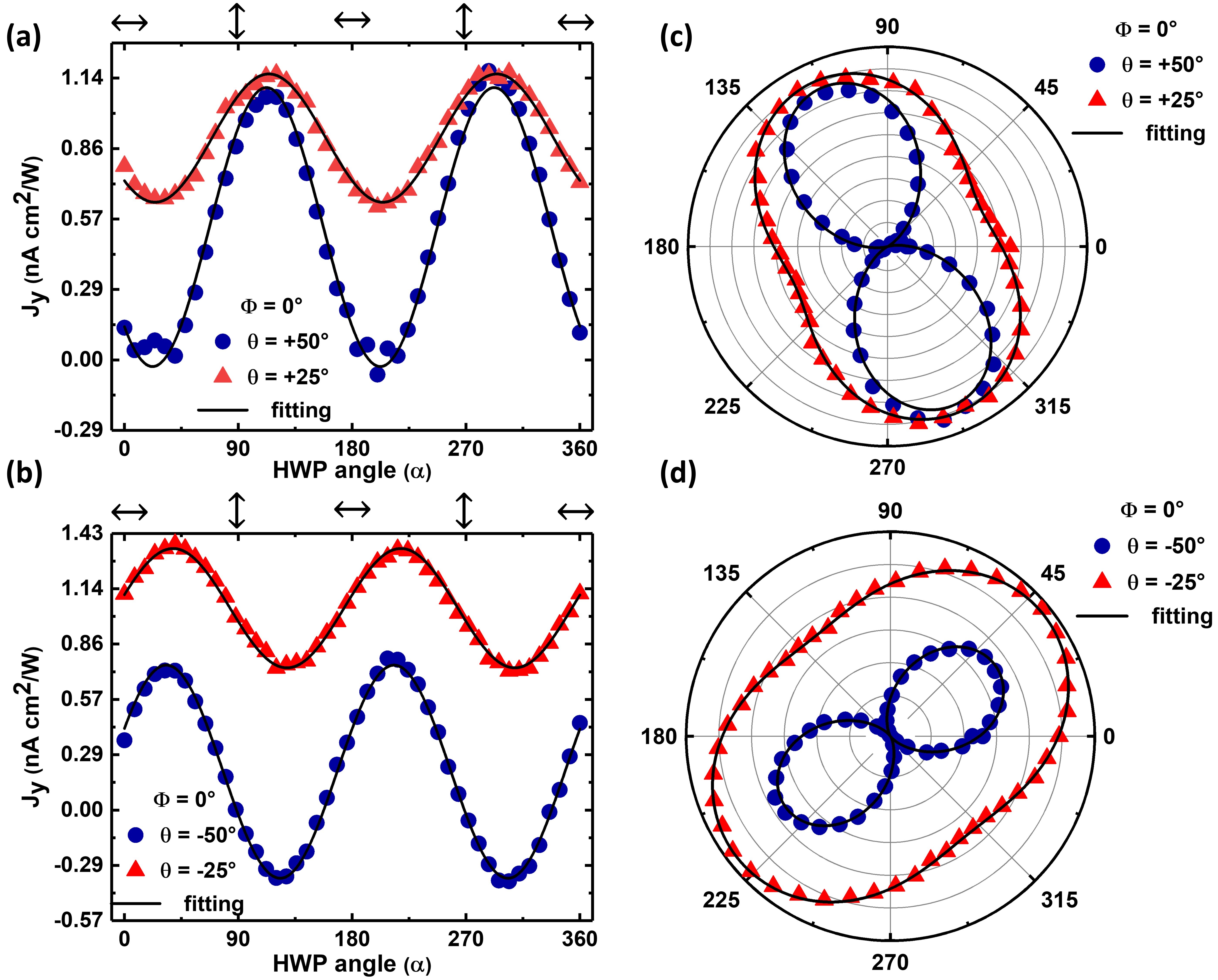}
\caption{\label{fig:5}(a,b) Photocurrent density $J_y$, normalized by the incident light intensity, as a function of the half wave plate rotation angle $\alpha$ for selected positive (a) and negative (b) angles of incidence $\theta$ relative to the sample normal. The corresponding polarization states of the incident light for different values of $\alpha$ are indicated on the top axis. (c,d) Polar plots of the same data shown in panels (a) and (b), respectively. Solid lines represent fits to Eq.~(2), as discussed in the main text.}
\end{figure*}

To ensure that the measured polarization dependent photocurrent arises from a second-order response, we examined its dependence on incident laser power by fitting  to Eq.(\ref{qwp}) at a fixed incidence angle of $\theta = -50^\circ$, as shown in Fig.\ref{fig:4}(a). All extracted components ($C$, $L_1$, $L_2$, and $D$) scale linearly with power over the range 0--40~mW, confirming operation within the linear regime, as shown in the Fig.\ref{fig:4}(b).

To further clarify the origin of the linear polarization dependent photocurrent, we performed polarization resolved measurements by rotating a half wave plate (HWP) for different incident angle $\theta$. This approach allows controlled modulation of the linear polarization orientation of the incident light without introducing helicity dependent components. The resulting photocurrent as a function of the HWP rotation angle is shown in Fig.\ref{fig:5}(a,b) and corresponding polar plots in Fig.\ref{fig:5}(c,d). The experimental data are well described by the fitting function $j(\theta)=A\sin(2\theta)+B\cos(2\theta)$, indicating that the observed response is dominated by mechanisms sensitive to linear polarization.

\begin{figure*}
\centering
\includegraphics[width=\textwidth]{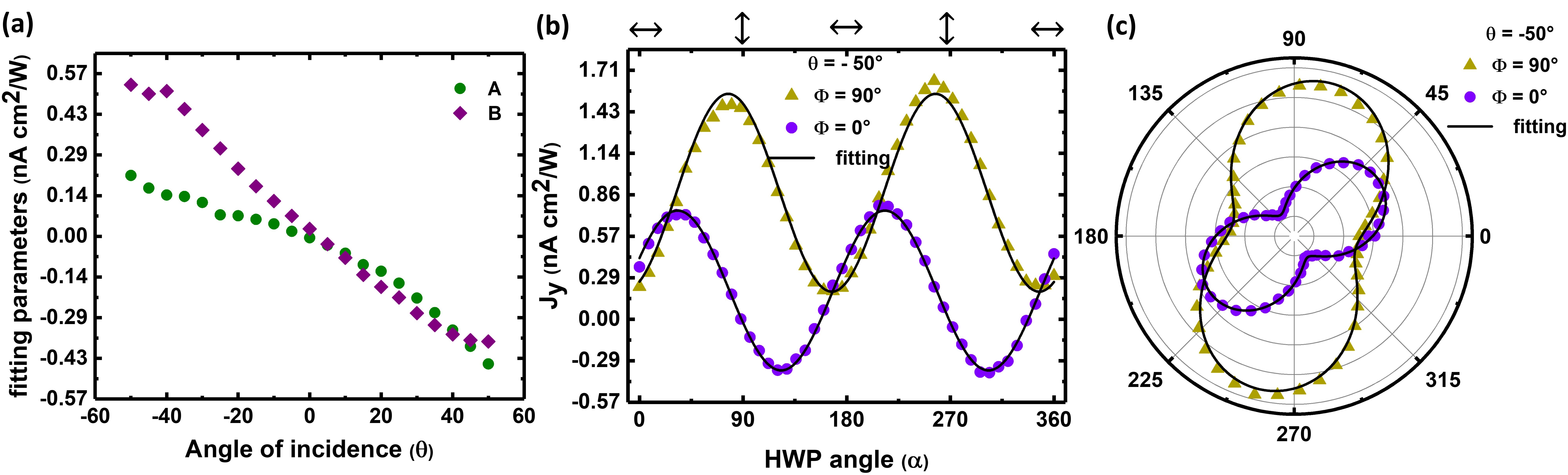}
\caption{\label{fig:6}(a) Dependence of the fitting parameters $A$ and $B$ on the angle of incidence $\theta$. (b) Polarization-dependent modulation of the photocurrent measured at different values of $\phi$ for a fixed angle of incidence $\theta = -50^\circ$, usinh a HWP. (c) Polar representation of the same data shown in panel (b).}
\end{figure*}

The extracted coefficients $A$ and $B$ exhibit a strong dependence on the incident angle $\theta$ Fig.\ref{fig:6}(a), closely mirroring the angular behavior of the linear polarization components $L_1$ and $L_2$ obtained from the QWP measurements discussed earlier. Specifically, both coefficients are strongly suppressed at normal incidence ($\theta=0$) and increase in magnitude with increasing $\theta$, while reversing sign for opposite incidence angles. This odd parity with respect to $\theta$ reflects the dependence of the photocurrent on the in-plane component of the photon wave vector, which vanishes at normal incidence and increases as $\sin\theta$ under oblique illumination. As a result, the linear polarization dependent photocurrent attains a minimum at normal incidence. The observed photocurrent as a function of the HWP rotation angle for different planes of incidence is shown in Fig.\ref{fig:6}(b) and the corresponding polar plots in Fig.\ref{fig:6}(c).

The close correspondence between the $\theta$ dependences of $A$, $B$ and those of $L_1$, $L_2$ demonstrates that the HWP and QWP based analyses probe the same underlying physical mechanisms, albeit in different polarization bases. While the QWP measurements enable separation of helicity dependent and helicity independent contributions, the HWP configuration isolates the purely linear polarization response. The consistency between these two approaches supports a photon drag mediated origin of the linear polarization dependent photocurrent, arising from momentum transfer from the electromagnetic field to charge carriers. These observations provide compelling evidence that the linear photocurrent components are governed by symmetry allowed photon drag processes requiring finite in-plane photon momentum. Recent terahertz emission studies in acentrosymmetric Dirac semimetals have revealed a giant photon momentum locked response, in which the direction and amplitude of the emitted THz radiation are governed by the incident photon momentum, establishing photon drag as a dominant nonequilibrium mechanism in topological semimetals \cite{cheng2023giant}. Recent observations of the nonlinear Hall effect in bilayer $\mathrm{WTe_2}$ under time-reversal symmetric conditions demonstrate how a finite Berry curvature dipole, arising from intrinsic inversion symmetry breaking, gives rise to second-order responses including helicity dependent photocurrents and nonlinear transverse voltages \cite{ma2019observation}. In such noncentrosymmetric systems, the broken parity of the crystal potential permits a nonvanishing Berry curvature dipole, enabling second-order responses even in the absence of magnetic fields. In contrast, the present results demonstrate that analogous quantum geometric responses can be realized in a centrosymmetric bulk system. Here, the helicity dependent photocurrent is not an intrinsic bulk property but instead emerges from finite photon momentum, which acts as an effective symmetry breaking axis. By going beyond the electric-dipole approximation and incorporating boundary induced symmetry breaking, our observations establish that nonlinear transport governed by Berry curvature can be accessed in a broader class of materials without requiring intrinsic bulk inversion asymmetry. These results demonstrate that momentum transfer from light can efficiently drive directional charge dynamics beyond conventional photogalvanic effects. Our observations extend this concept to centrosymmetric PdTe, where photon momentum similarly provides an effective symmetry breaking axis, enabling polarization dependent dc photocurrents in a system where such responses are nominally forbidden.

\section{\label{sec:level1}Conclusion}

In summary, our findings demonstrate that, in the centrosymmetric Dirac semimetal PdTe, photon momentum transfer acts as an effective dynamic symmetry breaking mechanism that activates second order nonlinear photocurrents otherwise forbidden by crystal symmetry. Through polarization resolved measurements, we disentangle distinct contributions from the circular photogalvanic effect (CPGE), geometric shift currents, and photon drag mediated processes. Importantly, the helicity dependent photocurrent exhibits a clear thickness dependence, evidencing a substantial bulk contribution arising from momentum assisted optical transitions rather than being confined to surface states. These observations provide direct experimental support for recent theoretical predictions linking such forbidden nonlinear responses to the underlying interband quantum geometry. Beyond fundamental insights, this work establishes photon momentum as a practical and tunable control parameter for generating directional dc currents in high symmetry materials without the need for structural asymmetry or external bias. Such capability offers a scalable route toward compact, polarization sensitive photodetectors, low power optical switches, and polarization sensitive energy conversion devices. Overall, PdTe and related centrosymmetric Dirac materials emerge as promising platforms for next generation optoelectronic technologies that leverage intrinsic quantum properties for functional device engineering.

\begin{acknowledgments}

The authors acknowledge financial support from the University Grants Commission (UGC) and the Ministry of Education (MoE), Government of India, through their research funding initiatives.

\end{acknowledgments}

\section*{Data Availability}

The data supporting the findings of this article are not
publicly available. The data are available from the authors
upon reasonable request.

\bibliography{apssamp}

\end{document}